# Integrated lithium niobate microwave photonic processing engine

Hanke Feng[1,7], Tong Ge[1,7], Xiaoqing Guo[1,2], Benshan Wang[3], Yiwen Zhang[1], Zhaoxi Chen[1], Sha Zhu[1,4], Ke Zhang[1], Wenzhao Sun[1,5,6], Chaoran Huang[3], Yixuan Yuan[1,3] and Cheng Wang[1*]

[1]*Department of Electrical Engineering & State Key Laboratory of Terahertz and Millimeter Waves, City University of Hong Kong, Kowloon, Hong Kong, China*
[2]*Department of Engineering Science, University of Oxford, Oxford, UK*
[3]*Department of Electronic Engineering, Chinese University of Hong Kong, Shatin, Hong Kong, China*
[4]*College of Microelectronics, Faculty of Information Technology, Beijing University of Technology, Beijing, China*
[5]*City University of Hong Kong (Dongguan), Dongguan, China*
[6]*Center of Information and Communication Technology, City University of Hong Kong Shenzhen Research Institute, Shenzhen, China*
[7]*These authors contributed equally*
*cwang257@cityu.edu.hk

**Abstract:** Integrated microwave photonics (MWP) is an intriguing field that leverages integrated photonic technologies for the generation, transmission, and manipulation of microwave signals in chip-scale optical systems [1,2]. In particular, ultrafast processing and computation of analog electronic signals in the optical domain with high fidelity and low latency could enable a variety of applications such as MWP filters [3-5], microwave signal processing [6-9], and image recognition [10,11]. An ideal photonic platform for achieving these integrated MWP processing tasks shall simultaneously offer an efficient, linear and high-speed electro-optic (EO) modulation block to faithfully perform microwave-optic conversion at low power, and a low-loss functional photonic network that can be configured for a variety of signal processing tasks, as well as large-scale, low-cost manufacturability to monolithically integrate the two building blocks on the same chip. In this work, we demonstrate such an integrated MWP processing engine based on a thin-film lithium niobate (LN) platform capable of performing multi-purpose processing and computation tasks of analog signals up to 256 giga samples per second (GSa/s) at CMOS-compatible voltages. By integrating a high-speed EO modulation block and a multi-purpose low-loss signal processing section on the same chip fabricated from a 4-inch wafer-scale process, we demonstrate high-speed analog computation, i.e., first- and second-order temporal integration and differentiation with processing bandwidths up to 67 GHz and computation accuracies up to 98.0 %, and deploy these functions to showcase three proof-of-concept applications, namely, ordinary differential equation (ODE) solving, ultra-wideband (UWB) signal generation and high-speed edge detection of images. We further leverage the image edge detector to enable a photonic-assisted image segmentation model that could effectively outline the boundaries of melanoma lesion in medical diagnostic images, achieving orders of magnitude faster processing speed and lower energy consumption than conventional electronic processors. Our ultrafast LN MWP engine could provide compact, low-power, low-latency, and cost-effective solutions for future wireless communications, Internet of things, high-resolution radar systems and photonic artificial intelligence.

**Main**

The rapid expansion of wireless networks, Internet of Things (IoTs), and cloud-based services is posing pressing challenges on the electronic bandwidth, processing speed, and power consumption of underlying radio frequency (RF) systems [1,2]. The burgeoning artificial intelligence (AI) technologies also demand ultrahigh-speed, low-latency, and low-power processing and computation of analog signals much beyond those offered by traditional electronic integrated circuits. Microwave photonics (MWP) technology provides effective solutions to address these challenges through the usage of optical components to perform microwave signal generation, transmission and manipulation tasks [1]. Its wide operation bandwidth and low loss characteristics also allow ultrahigh-speed and long-distance signals analysis and detection missions [1]. Recently, the surge of photonic integration technologies has further led to a dramatic reduction in the size, weight, and power (SWaP) of MWP systems with enhanced robustness and functionalities, termed integrated MWP [2]. Impressive demonstrations of integrated MWP applications include arbitrary RF waveform generation [13,14], true-time delay beamforming [15], instantaneous frequency measurement [16] and so on.

Despite the tremendous progress, integrated MWP systems still face substantial challenges in performing ultrahigh-speed analog signal processing tasks with chip-scale integration, high fidelity, and low power at the same time. An ideal photonic platform to meet these demands should support electro-optic (EO) modulators with low drive voltages, broad bandwidths and high linearity to faithfully convert microwave signals into optical signals, as well as a versatile low-loss functional device toolbox for further processing the converted signals in the optical domain. To date, most MWP applications have been demonstrated in silicon photonics [7,8,13,14,16] owing to its low-cost, large-scale fabrication readiness and the wide availability of functional devices. However, the free carrier-based modulation mechanism [17] in silicon is intrinsically accompanied with nonlinear EO response, large carrier-absorption loss, limited response speed, and unsatisfactory power-handling ability, resulting in critical trade-offs between the signal fidelity, power consumption, operation bandwidth and signal-to-noise ratio achievable in silicon MWP systems. Indium phosphide (InP) is another attractive platform for MWP systems with potential for monolithic integration of active and passive photonic elements on the same chip [4,6]. However, the relatively large propagation loss and small index contrast in InP waveguides [2], together with yield issues, significantly hurdle the performances and functionalities achievable in the signal processing section of future large-scale MWP systems. While silicon nitride (SiN) [18,19] is an excellent MWP platform by virtue of its ultra-low propagation loss and high-power handling ability, the lack of second-order nonlinearity prevents the realization of high-speed EO modulators in a monolithic SiN platform.

As a result of these material trade-offs, many integrated MWP systems have been realized by combining silicon or SiN photonic chips with traditional off-the-shelf lithium niobate (LN) modulators. This approach has enabled advanced computation and information processing tasks, such as differentiation, integration and Hilbert transformation [20-26], at the expense of increased bulkiness, system complexity and power consumption. Such signal processing and computation have also been demonstrated in all-optical circuits [6,27,28], offering potential processing bandwidths up to several THz [28]. However, the processed signals in these demonstrations are usually limited to simple Gaussian or Gaussian-derived waveforms generated by contemporary all-optical techniques, e.g., mode-locked lasers, while signals urgently in need of high-speed processing capability today are often much more complicated and arbitrary and can only be accessed from the electronic domain. Another potential solution is heterogeneous technology [16,29-31] that integrates III-V lasers/photodetectors and silicon modulators with low-loss SiN passive photonics via hetero-epitaxial growth or wafer bonding, while featuring additional cost and complexity in fabrication.

The recently emerged thin-film lithium niobate platform is a promising candidate to address these urgent demands and critical challenges by integrating efficient EO convertors and low-loss signal processors on the same chip [32]. The Pockels effect in LN is intrinsically linear, instantaneous and low-loss, ideally suited for realizing high-fidelity microwave-optic signal conversion with low power consumption and broad bandwidths [33]. Recently, many miniaturized and high-performance LN modulators have been demonstrated, exhibiting bandwidths covering the entire microwave and millimeter-wave bands [34-36], CMOS-compatible drive voltages [37,38], and ultra-high linearity performance [39]. Moreover, a full range of high-performance and low-loss functional devices are now endowed on the same platform, including ultrahigh-$Q$ microresonators [40], programmable filters [41], efficient frequency converters/shifters [42] as well as low-loss delay lines [43]. Efforts to scale up these elements into LN photonic integrated circuits (PIC) with low propagation loss and wafer-scale manufacturability have recently further boosted the cost effectiveness and commercial relevance towards a potentially high-performance, large-scale, and multi-purpose LN MWP system [44,45].

Here, we fulfill this promise by demonstrating a high-fidelity, broadband and low-power-consumption MWP system leveraging a 4-inch wafer-scale LN platform, realizing high-speed analog computation of microwave signals up to 256 GSa/s with multi-purpose functionalities, i.e., first- and second-order temporal integration and differentiation. Building upon these computation functions, we show three proof-of-concept applications, including ordinary differential equation (ODE) solving, ultra-wide bandwidth (UWB) signal generation, and high-speed edge-feature detection of images. Furthermore, we plug the photonics-assisted image-edge detector, with orders of magnitude higher computing speed and lower energy consumption than traditional electronics-based image-edge detection algorithms, into a neural network-based image segmentation model [46] and showcase the effective identification of melanoma lesion outlines in medical diagnostic images.

**Results**

Figure 1a shows the schematic illustration of our multi-purpose MWP system on LN platform, consisting of an efficient EO modulation section for transferring high-speed microwave signals into optical domain, and a low-loss signal processing section for realizing the analog computation functions, including first- and second-order temporal integration and differentiation. The PICs are directly patterned on a 4-inch thin-film LN wafer (Fig. 1b) using an ultraviolet (UV) stepper lithography system (see Methods). Our LN MWP platform supports a variety of high-performance device building blocks, including microring resonators with ultrahigh intrinsic quality ($Q$) factors up to 6 million (corresponding to propagation loss ~ 5 dB/m), low-voltage and broad bandwidth (> 67 GHz) intensity/phase modulators with advanced slotted-electrodes [47], add-drop microring resonators as temporal integrators, unbalanced Mach-Zehnder interferometers (MZIs) as differentiators, as well as respective cascaded versions for second-order integration and differentiation tasks. The corresponding microscope images of these photonic building blocks and their measured critical performance metrics are shown in the Fig. 1c. False-colored scanning electron microscope images shown in Fig. 1d highlight details of the waveguide sidewall, the coupling region of a microring, the cross-section of a waveguide, and a multi-mode interference (MMI) coupler, respectively. Based on these MWP building blocks, we next discuss two ultrahigh-speed microwave signal computation processes, i.e. temporal integration and differentiation, as well as their applications.

**High-speed microwave photonic temporal integrator**

Figure 2a illustrates the working principle of our high-speed MWP temporal integrator, which consists of an MZI intensity modulator and add-drop microring resonators, and is designed to take the temporal integration of a complex input microwave signal and output in the form of optical intensity (see Methods). An integration task in the time domain is equivalent to a frequency response of the system $H(\omega) = \frac{1}{j(\omega-\omega_0)}$ in the frequency domain [27], where $j = \sqrt{-1}$, $\omega$ is the optical angular frequency and $\omega_0$ is the carrier frequency of the signal to be processed. Here we adopt add-drop microring resonators to serve as integrators [6,22,27], where the Lorentzian lineshape approximately follows the above frequency response within the resonance bandwidth, as shown in Fig. 2c.

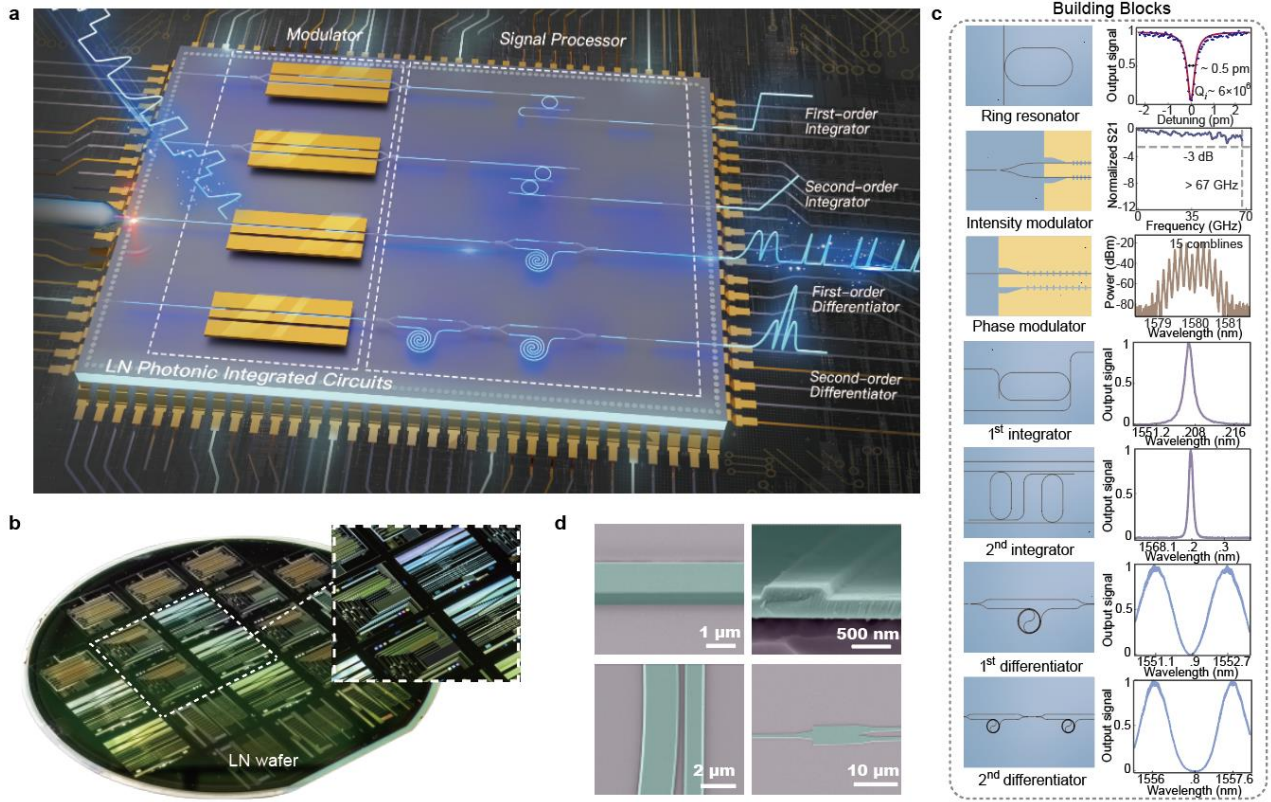

**Fig. 1. Wafer-scale LN-based MWP signal processing engine and its building blocks. a** Schematic diagram of LN-based MWP processing engine, consisting of a high-speed EO modulation section that faithfully converts analog electronic signals into optical domain, and a low-loss multi-purpose photonic processing section. **b** Photography of 4-inch wafer-scale LN photonic integrated circuits patterned using a UV stepper lithography system. **c** Microscope images and key performance metrics of the fundamental building blocks of our high-speed MWP systems, including micro-resonators with intrinsic quality factor $\sim 6 \times 10^6$, low-driving-voltage and broad bandwidth intensity and phase modulators for signal encoding, add-drop ring resonators as integrators, unbalanced MZIs as differentiators, as well as cascaded rings and MZIs as second-order integrators and differentiators. **d** False-color scanning electron micrographs (SEM) of the devices, showing the sidewall of a waveguide, the coupling region of a micro-resonator, the cross-section view of a waveguide and a multi-mode interference (MMI) coupler, respectively.

This first-order integrator has a measured free spectral range (FSR) of 80 GHz, a loaded $Q$ factor of $\sim 0.9$ million measured at the drop port, which corresponds to a photon lifetime of 700 ps (Fig. 1c). We characterize the performance of our LN-based MWP integrator using the experimental setup shown in Fig. 2b. Notably, the measurements are performed using small signals (peak voltage = 500 mV) directly generated from an arbitrary-waveform generator (AWG) without the use of microwave amplifiers, enabled by the low-driving voltage and high signal-to-noise ratio of our LN modulators, promising for future low-power-consumption and high-fidelity MWP systems. The integration results in the form of optical intensity are recorded by a high-speed photodetector (PD) through homodyne detection with the optical carrier itself when the modulator is biased at the quadrature point. (see Methods)

We first test the basic performance of our first-order integrator by injecting a Gaussian pulse with a full width at half maximum (FWHM) of 90 ps. The output signals in Fig. 2d (*i*) clearly show a step-like waveform with an integration time up to ~ 600 ps (defined as the decay time to reach 80 % of the maximum intensity [27]). Using resonators with even higher *Q*-factors could further increase the integration time, at the expense of lower throughput due to narrower resonance linewidths (or operation bandwidths) [6]. Next, we demonstrate coherent integration of more advanced waveforms by injecting in-phase (*ii*) and out-of-phase (*iii*) doublet pulses, as well as triplet pulses (*iv*), the corresponding results of which are shown in Fig. 2d. Specifically, the results for the in-phase doublet pulses (*ii*) feature a clear double-step function due to the constructive addition of the two waveforms, and the duration time of the step profile matches the interval between the two pulses. Such functions could potentially find applications in high-speed bit counting. In contrast, when the two pulses are out of phase (*iii*), the time integral of the second optical pulse cancels with that of the first one, leading to a rapid return to the baseline. Different duration time can be obtained by setting the position of the second pulse to realize signal memory functions [48]. We then input a more complex triplet pulse (*iv*) to verify the capability of processing complicated RF signals, the result of which shows an average accuracy of 95.9 % compared with the ideal output. Based on the results of the first-order temporal integrator, we further demonstrate second-order integration function by cascading two microring resonators with aligned resonance peaks (Fig. 2c) (see Methods). When injecting a single Gaussian pulse (FWHM ~ 120 ps), the corresponding output (Fig. 2d, *v*) matches well with the ideal prediction, with an average accuracy of 95.8 % within the device lifetime, and shows a much longer rise time than the duration of the input pulse, indicating the effective realization of second-order integration function.

We then apply this integrator to showcase our first MWP application, i.e. solving ordinary differential equations (ODE) for high-speed electronic signals, expressed as: $\frac{dy(t)}{dt} + ky(t) = x(t)$, where $x(t)$ represents the input RF signal, $y(t)$ is the solution to be determined, and $k$ represents an arbitrary constant [22]. This basic ODE can be utilized to model a broad range of basic engineering systems and physical phenomena, such as temperature diffusion processes and automatic control systems [22,49]. The process of solving this ODE in the time domain could be described canonically as an integration function embedded inside a feedback loop, which is equivalent to a Lorentzian frequency response of $H(\omega) = \frac{1}{j(\omega-\omega_0)+k}$ (Fig. 2e). We achieve this Lorentzian frequency response again using an add-drop ring resonator, where the loaded $Q$ factor determines the constant $k$ in the ODE [22](see Methods). As a proof of concept, we fabricate and test three different ring resonators with $Q$ factors of $4.8\times10^5$, $1.7\times10^5$, and $0.8\times10^5$ by controlling the coupling gaps (Fig. 2f), corresponding to $k$ values of 1.24 ns$^{-1}$, 3.49 ns$^{-1}$ and 7.44 ns$^{-1}$, respectively. During the testing, we use 400 ps super-Gaussian pulses as input signals to better distinguish the solutions from the input [49]. The respective results with different $k$ constants in Fig. 2g show excellent agreement with simulated solutions (dashed lines) obtained by an ideal ODE solver, with an average computation accuracy of 98.1 %. Compared with algorithms in traditional electronics that require multiple iterations, our MWP system solves the ODE almost instantaneously as photons pass through, significantly improving the processing speed while maintaining excellent computation accuracy. Further equipping the resonators with thermo-optic or EO tunable couplers to actively control the loaded $Q$ factor could enable a tunable ODE solver with variable coefficients to address more complex practical application scenarios [50].

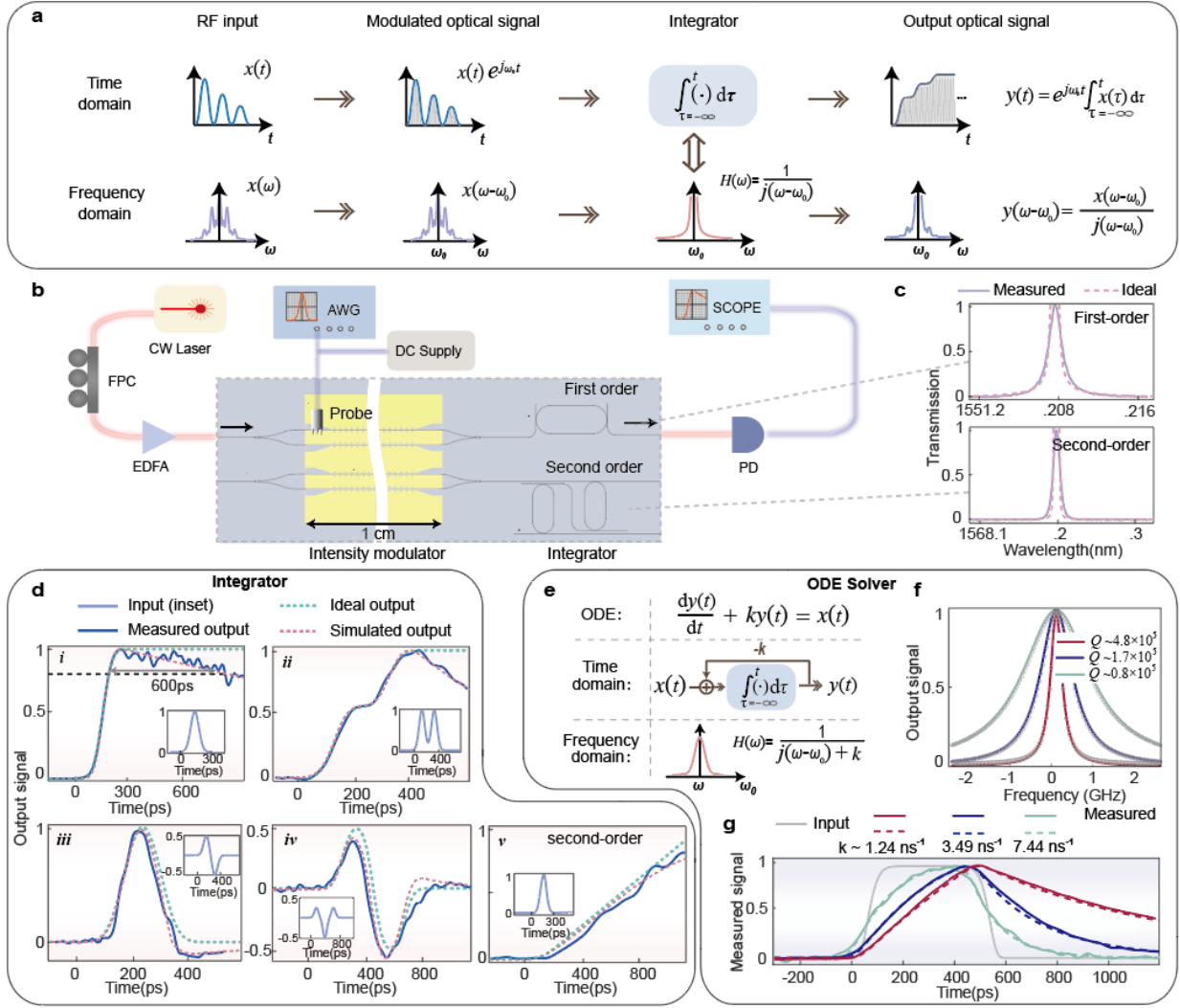

**Fig. 2. High-speed microwave photonic temporal integrator. a.** Working principle of our MWP temporal integrator. An integration task in the time domain is equivalent to a frequency response of $H(\omega) = 1/j(\omega - \omega_0)$. **b.** Experimental setup for the measurement of the MWP integrator. Inset shows a microscope image of the device. AWG, arbitrary waveform generator; FPC, fiber polarization controller; EDFA, erbium-doped fiber amplifier; PD, photodetector. **c.** Measured (solid) and ideal (dashed) transmission for the first- and second-order integrators. **d.** Normalized measured temporal responses of the integrator (blue solid), together with simulated (red dashed) and ideal (green dashed) responses, for (*i*) first-order integration of a single Gaussian pulse showing an integration time of 600 ps, (*ii*) first-order integration of an in-phase doublet pulse featuring a clear double-step profile, (*iii*) first-order integration of an out-of-phase doublet pulse, showing a rapid return to the baseline when the second pulse comes in, (*iv*) first-order integration of a triplet pulse, and (*v*) second-order integration of a single Gaussian pulse. Insets: the corresponding input RF waveforms. **e.** Working principle of the ordinary differential equation (ODE) solving system. **f.** Measured frequency response of the ODE solvers with different *Q* factors corresponding to three different *k* coefficients. **g.** Normalized measured (solid) and simulated (dashed) ODE solutions of a 400-ps input super-Gaussian pulse (grey) for different *k* coefficients.

**High-speed microwave photonic temporal differentiator**

The second demonstrated high-speed MWP computation task is temporal differentiation, which takes the derivative of the input microwave signal and output in the form of optical field or intensity. Here, we deploy a frequency-chirp-based differentiation scheme [20] leveraging an EO phase modulator and an MZI-based differentiator on the LN platform (see Methods). The basic working principle is shown in Fig. 3a: the input RF signal $x(t)$ is first loaded on a continuous-wave optical signal by the phase modulator, leading to an instantaneous optical phase of $\omega_0 t + \beta x(t)$, where $\omega_0$ is the carrier frequency of the signal and $\beta$ is the modulation index. This induces an instantaneous frequency chirp of $\omega_0 + \beta \frac{dx(t)}{dt}$ that exactly follows the differentiation of the input signal $\frac{dx(t)}{dt}$. The chirped frequency information is then mapped into optical field or intensity using a signal processing unit, i.e. an unbalanced MZI, which is carefully designed to provide the desired frequency response. Specifically, we achieve field-to-field (field-to-intensity) differentiation by biasing the MZI at the null (quadrature) point, where the output optical field (intensity) is linearly proportional to the optical frequency [20]. The top panel of Fig. 3c shows the measured optical transmission spectrum of the fabricated unbalanced MZI as a first-order differentiator, consistent with the intended linear frequency response within a processing bandwidth of 85 GHz, limited by FSR. Even higher processing bandwidths could be achieved using MZIs with larger FSRs, however, at the expense of lowered differentiation efficiency (determined by the spectral slope of the frequency response) [20]. We could also achieve second-order differentiation by cascading two unbalanced MZIs with aligned null wavelengths, the frequency response of which is shown in the bottom panel of Fig. 3c with a faithful processing bandwidth of 65 GHz.

We test the basic field-to-field/intensity differentiation performance by injecting a sequence of RF signals including Gaussian pulses, square pulses and stepped pulses [Fig. 3d (blue)] into the phase modulator, using the experimental setup shown in Fig. 3b. All the demonstrations are carried out in the small-signal regime. The red trace in Fig. 3d shows the corresponding measured field-to-field differentiation results when the MZI is biased at the null point, where the pulse height is determined by the temporal rising/falling slope of input signals. Here the output signals are positive for both rising and falling edges since the differentiation result in the form of optical field is measured by a direct intensity detection at the PD showing $\left|\frac{dx(t)}{dt}\right|^2$. In contrast, the field-to-intensity differentiation results (yellow trace in Fig. 3d) are carried by optical intensity and measured through a homodyne detection similar to that used in field-to-intensity integration experiments, showing positive (negative) pulses at rising (falling) slope of input signals. The right panels show blow-up views of the output waveforms, exhibiting good agreement with the simulation ones.

To showcase the unique ultrahigh-speed signal processing capability of our LN MWP system, we inject ultra-short Sinc pulses with a main-lobe FWHM duration of ~ 9.6 ps (corresponding to an analog bandwidth of ~ 62 GHz, limited by the bandwidth of our oscilloscope, as shown in the top right inset of Fig. 3e) to the device and obtain field-to-intensity differentiation result (bottom left panel in Fig. 3e) consistent with simulation, with an average computation accuracy of 98.0 % (see Methods). The corresponding Fourier spectrum of the differentiated signal is clearly reshaped from the initial signal, as shown in bottom right panel in Fig. 3e. We further verify that our devices can support accurate differentiation operations at least up to 67 GHz, by injecting sawtooth signals at different repetition rates and monitoring the optical spectra of output differentiation signals. Ideally, the derivative of a sawtooth signal features infinite numbers of frequency components at integer multiples of the fundamental frequency. In our real system, the analog bandwidth of the AWG (70 GHz) could faithfully preserve the first three harmonics of a 20-GHz sawtooth signal, which corresponds to a three-step-like signal in the time domain (blue solid line in the top left panel of Fig. 3f). After passing through the differentiator, the directly recorded time-domain signal (bottom left panel) and the measured optical spectrum (right panel) both show good agreement with the simulated result when considering only the first three harmonics, i.e. 20, 40, and 60 GHz. At even higher frequencies where the third harmonic is beyond the bandwidth of our oscilloscope for a direct time-domain measurement, we could still infer the differentiation performance of our MWP processor by monitoring the output optical power of the third harmonic and comparing with the simulation results. The results confirm that our MWP temporal differentiator could reliably process input analog signals at frequencies up to 67 GHz, currently limited by other test equipment including AWG, RF probes and cables. In addition, we test the second-order field-to-field differentiation performance by injecting a 120-ps Gaussian pulse to the cascaded MZI device (Fig. 3g), which is also in line with the expected result.

One attractive MWP application based on the differentiator is the generation of UWB signals, an emerging wireless protocol with wide bandwidth (3.1-10.6 GHz) and low power spectral density (< 41.3 dBm/MHz) for short-range high-throughput wireless communications and sensor networks [51] (Fig. 3h). Here we demonstrate the generation of UWB carrier pulses by performing field-to-intensity differentiation of Gaussian monocycle pulses with FWHM ~ 100 ps (Fig. 3i). The differentiation shows an anti-symmetric doublet pulse with an average accuracy of 97.4% compared with the ideal response. The corresponding RF spectrum of the output UWB pulse is shown in Fig. 3j, featuring a center frequency of 6.1 GHz and a 10-dB bandwidth of 7.3 GHz, with a fractional bandwidth of 120 %, in accordance with the Federal Communications Commission (FCC) regulations [51]. The demonstrated UWB signal generation on our LN MWP platform could provide compact and cost-effective solutions for next-generation wireless communications and remote sensing systems with seamless compatibility with optical networks.

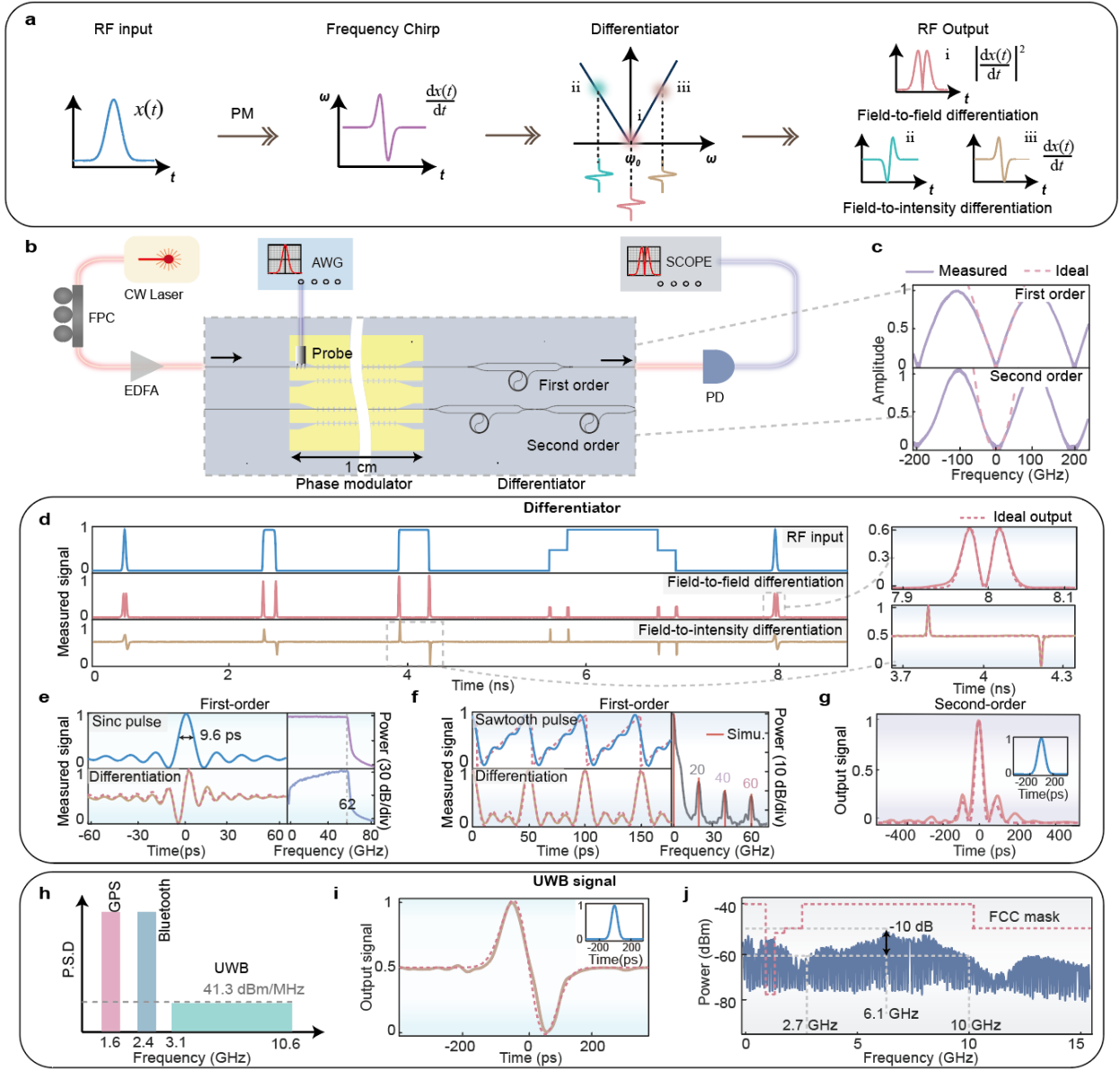

**Fig. 3. High-speed microwave photonic temporal differentiator. a.** Working principle of the frequency-chirp-based MWP temporal differentiator. Input RF signal $x(t)$ is first loaded on an optical carrier by a phase modulator to induce an instantaneous frequency chirp that follows the differentiation of the input signal. The chirped frequency information is then projected into optical field or intensity through an unbalanced MZI biased at the null or the quadrature point, respectively. **b.** Experimental setup for the measurement of MWP differentiator. Inset shows a microscope image of the device. **c.** Measured and ideal frequency response for the first- and second-order differentiators. **d.** Measured field-to-field differentiation response (red) and field-to-intensity differentiation results (yellow) for a sequence of input RF signals (blue). Blow-up panels show the measured and ideal results for a Gaussian pulse and a rectangle pulse. **e.** Measured field-to-intensity differentiation result (yellow solid), together with the ideal response (red dashed) for an ultrafast input Sinc pulse with FWHM of main lobe ~ 9.6 ps (blue). Right panel shows the corresponding Fourier spectra of the Sinc pulse (top) and its differentiation form (bottom) with an analog bandwidth of 62 GHz. **f.** Measured (yellow solid) and simulated (red dashed) field-to-intensity differentiation result (bottom left) for a 20-GHz sawtooth signal (top left). The actual input signal (blue) deviates from an ideal sawtooth (red dashed) since only the first three harmonics are preserved. Right panel shows the measured (black) and simulated (red) optical spectra of the differentiated signal, clearly resolving the first three sidebands. **g.** Second-order field-to-field differentiation results for a Gaussian input pulse as shown in the inset. **h.**

Illustration of the power spectrum density (P.S.D.) and frequency range of UWB signal compared with GPS and Bluetooth. **i.** Measured (yellow solid) and ideal (red dashed) UWB monocycle pulse generated by field-to-intensity differentiation of a Gaussian pulse (inset). **j.** Measured RF spectrum of the generated UWB signal from 0 to 15 GHz. Red dashed line shows the power spectrum mask regulated by Federal Communications Commission (FCC).

**High-speed photonic-assisted image segmentation system**

Finally, we show our high-speed LN photonic processing engine could enable applications beyond traditional MWP-related scenarios. As an example, we realize a photonic-assisted image edge detector for segmentation of medical images, which can provide quantitative analysis to help clinicians conduct accurate disease diagnosis and treatment. The extraction of image edge features is realized by performing field-to-field differentiation operations on a time-domain data stream serialized from 2D images (see Methods). We first showcase the power of our edge-feature detector by feeding the system with a 250×250-pixel 'CityU' logo, serialized as a 256 GSa/s data stream. The temporal differentiation and edge detection functions are performed "on-the-fly" within a short time ($t=250 \times 250 \times \frac{1}{256 \text{ GSa/s}}$ = 244 ns) and captured by a real-time oscilloscope. We de-multiplex the captured time-series data back into matrix format to form the reconstructed image, showing clearly resolved edge features with 96.6 % accuracy compared with the simulated results (Fig. 4a).

Importantly, the demonstrated image processor is three orders of magnitude faster and consumes lower energy than performing traditional algorithms in an electronic computer (see Methods). The processing speed is currently limited by the sampling rate of our AWG and could be further increased considering the large analog bandwidth of our EO modulators deep into the millimeter-wave band [34].

We then plug our high-speed photonic-assisted image edge detector into a deep convolution neural network (DCNN)-based image segmentation model for outlining the boundaries of melanoma lesion in medical diagnostic images with superior processing speed, energy consumption and accuracy. When processing complex and often low-contrast medical images, the fuzzy boundaries between abnormal and normal regions could lead to predictions with compromised accuracies. This situation could be substantially improved by feeding the DCNN with edge-detected information instead of original images, which can be integrated into an arbitrary encoder-decoder architecture in an end-to-end way for medical image segmentation process [46]. Figure 4b-c illustrates the flow diagram and working principle of the proposed edge-enhanced DCNN segmentation model that intakes raw RGB images and outputs segmentation results (see Methods). To optimize the segmentation model, we first train the model with concatenated dermoscope images and the corresponding melanoma edge information derived from simulated differentiation, emphasizing the representations around melanoma lesion boundaries. Based on the optimized model, the test dermoscope images are concatenated with experimentally extracted lesion edge information to generate the melanoma region prediction. Figure 4d shows the original melanoma lesion images captured from dermoscope, simulated and experimentally measured edge features, as well as the lesion regions segmented by our model, respectively. The tested average segmentation accuracy of our edge-facilitated model is 97.3 %, proving the effectiveness of the proposed photonic-assisted segmentation model. Most importantly, the demonstrated LN photonic-assisted image edge detector features much higher computation speed and lower energy consumption compared with other photonic platforms as well as traditional electronics (see Methods), which will pave the path for high-complexity, high-throughput and real-time medical diagnosis tasks. The functional toolbox of our LN photonic image processing engine could be further expanded and parallelized leveraging the excellent scalability of our platform, leading to more advanced functions like matrix multiplication [10] and enabling a variety of future photonics-enabled AI and computer vision technologies.

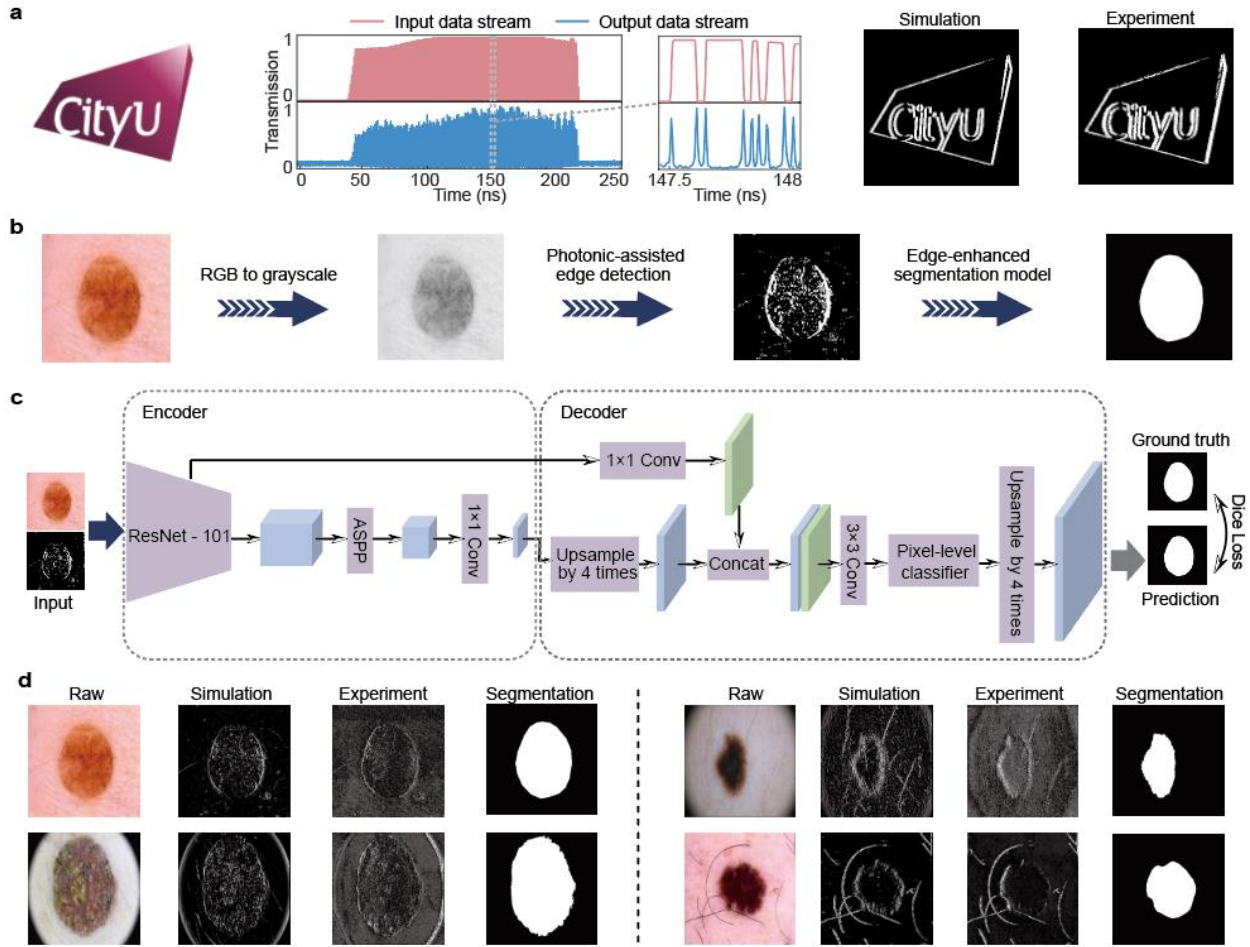

**Fig. 4. High-speed photonic-assisted medical image segmentation. a.** Photonic-assisted image edge detection of a 250 ×250-pixel 'CityU' logo. **b.** Flow diagram of the photonic-assisted image segmentation model. The edge features of melanoma legion images are extracted by our LN-based image-edge detector and passed through a DCNN-based segmentation model to obtain the lesion segmentation results. **c.** Schematic diagram of the image segmentation model. **d.** Examples of melanoma lesion segmentation results showing original dermoscope images, simulated and experimentally extracted edge features, as well as the lesion regions segmented by our model, respectively.

**Discussions**

In summary, we have designed, fabricated and demonstrated a high-fidelity and low-power-consumption integrated microwave signal processing system that performs temporal integration and differentiation operations for ultrahigh-speed electronic signals up to 256 GSa/s, enabling a variety of applications including ODE solving, UWB signal generation and edge detection of images. In particular, our photonic-assisted image segmentation model proves to be more powerful for disease diagnosis tasks, bringing exciting opportunities for intersections between integrated MWP technology and medical diagnosis. Our LN-based MWP system exhibits significant and all-around performance edges in terms of operation bandwidth, processing speed and energy consumption compared with other on-chip and off-chip MWP platforms, proving an efficient and cost-effective engine for future general-purpose analog signal processing and computing tasks.

The performance of the current system can be further improved, by utilizing micro resonators with even higher $Q$ factors [40] to prolong the integration time to nanoseconds [6], and MZIs with larger FSR to further increase the bandwidth of differentiation operation. The reconfigurability and programmability of our LN-based on-chip MWP signal processing system could be further enhanced by equipping thermo-optic or EO tuning sections on the processing blocks to allow real-time switching among processing functions and active tuning of the processing bandwidth for task-oriented applications [6,7,9]. Adoption of multiple input/output ports with controllable (de-)multiplexers could enable parallel linear processing [52] and space-division multiplexing [53]. Importantly, our demonstrated MWP system is highly compatible with other high-performance photonic components available on the integrated LN platform, such as microcombs [54], frequency shifters [42], and true delay lines [43], which could be further integrated towards even more advanced MWP functionalities. Meanwhile, we expect other components of the MWP system, such as low-noise laser sources, high-power handling PDs, electronic integrated circuits, and microwave amplifiers, could be assembled on the LN platform through heterogeneous integration schemes, leading to highly compact, cost-effective and high-performance integrated MWP systems for next-generation communications and information technologies.

## Methods

### Design and fabrication of the devices:

Devices are fabricated from a commercially available x-cut LNOI wafer (NANOLN), with a 500-nm LN thin film, a 2-μm buried $SiO_2$ layer, and a 500-μm silicon substrate. $SiO_2$ is first deposited on the surface of a 4-inch LNOI wafer as etching hard mask using plasma-enhanced chemical vapor deposition (PECVD). Optical waveguides, MZIs and microring resonators are then patterned on the entire wafer using an ASML UV Stepper lithography system (NFF, HKUST) die by die (1.5 cm×1.5 cm) with a resolution of 500 nm. Next, the exposed resist patterns are transferred first to the $SiO_2$ layer using a standard fluorine-based dry etching process, and then to the LN device layer using an optimized Ar+ based inductively-coupled plasma (ICP) reactive-ion etching process. The LN etch depth is ~ 250 nm, leaving a 250-nm-thick slab. After removal of the residual $SiO_2$ mask and redeposition, an annealing process is carried out. Next, a second stepper lithography, metal evaporation and lift-off process are used to fabricate the microwave electrodes. The positive and negative electrodes are spaced by a gap of 5 μm to ensure strong EO coupling while minimizing metal-induced optical losses. Finally, chips are carefully cleaved for end-fire optical coupling with coupling loss ~ 4 dB per facet.

### Characterization of the building blocks for MWP signal processing:

More details are presented here for the LN building blocks shown in Fig. 1. For optical characterizations, a continuous-wave pump laser (Santec TSL-510) is sent to the devices under test using a lensed fiber after a polarization controller to ensure TE polarization. The output optical signal is collected using a second lensed fiber and sent to a 125-MHz PD (New Focus 1811) for low-frequency measurements. The optical transmission of the racetrack resonator (waveguide width ~ 2 μm) is fitted by a Lorentzian function, with a loaded $Q$ factor ~ 3 million in the critical coupling state ($Q_i$ ~ 6 million), indicating a corresponding propagation loss of ~ 5 dB/m.

For the on-chip MZI modulator, the small-signal EO $S_{21}$ response is measured by injecting small RF signals from a vector network analyzer (VNA, Keysight N5227B, 67 GHz) into the modulation electrodes via a high-speed probe (GGB industries, 67 GHz) and monitoring the output signals captured by a high-speed PD (Finisar XPDV412xR, 100 GHz) at various frequencies at the other port of the VNA. RF cable losses, probe loss and PD response are calibrated and de-embedded from the measured $S_{21}$ responses, showing 3-dB EO modulation bandwidths > 67 GHz. The modulator design includes tapered and micro-structured electrodes to fulfill the impedance- and velocity-matching conditions, simultaneously, while maintaining low RF loss [47]. The measured low-frequency half-wave voltage ($V_\pi$) is 2.6 V for a modulation length of 1 cm, corresponding to a voltage–length product of 2.6 V·cm.

For the on-chip phase modulator, an optical frequency comb consisting of 15 comb lines could be obtained by driving the modulator with a moderate 18-GHz microwave signal with power of 630 mW, corresponding to a total acquired phase shift of ∼ 1.05 π. The comb spectrum is measured by an optical spectrum analyzer (OSA, Yokogawa AQ6370).

To implement temporal integration, a symmetrically over-coupled add-drop racetrack resonator with a free spectral range (FSR) ~ 80 GHz is fabricated, with a loaded $Q$-factor ~ $9 \times 10^5$ and near-unity on-resonance transmission measured at the drop port. For second-order integration, two cascaded racetrack resonators with different FSRs (80 GHz and 82 GHz) are designed to align the two resonance peaks via Vernier effect. The envelop FSR is measured to be 26.24 nm, which is in line with the design value. For temporal differentiation, the asymmetrical MZI and cascaded MZIs are designed with FSR ~ 200 GHz to provide a reasonable balance between operation bandwidth and differentiation slope efficiency. The spiral waveguides are designed with a minimum radius of curvature of ~ 100 μm to minimize the footprint and limit excessive radiation loss.

### Principles of the microring-based MWP temporal integrator and ODE solver:

More details are presented here for the principles of microring-based temporal integrator and ODE solver. The integration results are represented in optical intensity changes on top of a DC intensity component when the amplitude modulator is biased at the quadrature point through a bias-tee, which are captured by an AC-coupled high-speed PD. The optical field right after amplitude modulation can be expressed as:

$$E_{in}(t) = E_0 e^{j\omega_0 t} \cos[\beta x(t) + \varphi_0] \quad (1)$$

where $\beta$ is the modulation index defined as $\pi V_p/V_\pi$, $V_p$ is the peak voltage applied to the driving electrodes, $E_0$ is the input electric field amplitude, $\varphi_0$ is the bias phase of the amplitude modulator, which is $-\pi/4$ here, $\omega_0$ is the optical carrier frequency and $x(t)$ is the normalized input microwave signal. Under small signal modulation ($\beta \ll 1$), the expression can be simplified as:

$$E_{in}(t) \approx \frac{\sqrt{2}}{2} E_0 e^{j\omega_0 t} [\beta x(t) + 1] \quad (2)$$

After Fourier transform, the input signal can be rewritten in the frequency domain as:

$$E_{in}(\omega) = \mathcal{F}\left[\frac{\sqrt{2}}{2}\beta E_0 e^{j\omega_0 t} x(t)\right] + \sqrt{2}\pi E_0 \delta(\omega - \omega_0) \quad (3)$$

After going through the integrator/ODE solver with a frequency response of $H(\omega) = \frac{1}{j(\omega-\omega_0)+k}$, the output signal can be written as:

$$E_{out}(\omega) = E_{in}(\omega)H(\omega) = \frac{\mathcal{F}\left[\frac{\sqrt{2}}{2}\beta E_0 e^{j\omega_0 t} x(t)\right]}{j(\omega-\omega_0)+k} + \frac{\sqrt{2}\pi E_0 \delta(\omega-\omega_0)}{j(\omega-\omega_0)+k} \quad (4)$$

When performing integration operations, we use microring resonators with ultrahigh Q-factors ($k \approx 0$), such that the output signal after inverse Fourier transform back to the time domain is:

$$E_{out}(t) = \mathcal{F}^{-1}[E_{out}(\omega)] \approx \mathcal{F}^{-1}\{\frac{\mathcal{F}\left[\frac{\sqrt{2}}{2}\beta E_0 e^{j\omega_0 t} x(t)\right]}{j(\omega-\omega_0)} + \frac{\frac{\sqrt{2}}{2} E_0 \delta(\omega-\omega_0)}{k}\}$$

$$= \frac{\sqrt{2}}{2}\beta E_0 e^{j\omega_0 t} \int x(t) dt + \frac{\sqrt{2}}{2} e^{j\omega_0 t} \frac{E_0}{k} \quad (5)$$

The output current of the PD could then be written as:

$$I_{out}(t) \propto \Re|E_{out}(t)|^2 = \frac{\Re E_0^2}{2k^2} + \frac{1}{2}\Re\beta^2 E_0^2 \left[\int x(t)dt\right]^2 + \frac{\Re\beta E_0^2}{k}\int x(t)dt$$

$$\approx \frac{\Re\beta E_0^2}{k}\int x(t)dt \quad (6)$$

where $\Re$ is the responsivity of the PD. Here the AC-coupled PD only responds to the third term in Eq. (6) since the first term is time-invariant and the second term is small for small signals. Therefore, the output signal of the PD directly corresponds to the integration result [$I_{out}(t) \propto \int x(t)dt$], which is generally the case for all field-to-intensity integration and differentiation results in this work.

On the other hand, when the internal decay loss of the ring resonator is non-negligible ($k \neq 0$), the output signal can be rewritten as:

$$E_{out}(t) = \mathcal{F}^{-1}[E_{out}(\omega)] \approx \mathcal{F}^{-1}\{\frac{\mathcal{F}\left[\frac{\sqrt{2}}{2}\beta E_0 e^{j\omega_0 t} x(t)\right]}{j(\omega-\omega_0)+k} + \frac{\frac{\sqrt{2}}{2} E_0 \delta(\omega-\omega_0)}{k}\}$$

$$= \frac{\sqrt{2}}{2}\beta E_0 e^{j\omega_0 t} e^{-kt}\int e^{kt} x(t)dt + \frac{\sqrt{2}}{2} e^{j\omega_0 t} \frac{E_0}{k} \quad (7)$$

The output PD signal is then:

$$I_{out}(t) \propto \Re|E_{out}(t)|^2 \approx \frac{\Re\beta E_0^2}{k} e^{-kt}\int e^{kt} x(t)dt \propto e^{-kt}\int e^{kt} x(t)dt \quad (8)$$

which is the solution of the first-order ODE solved in this work.

**Principles of the MZI-based MWP temporal differentiators:**
More details are presented here for the temporal differentiator. After loading the electronic signal into LN phase modulator, the electric field can be expressed as:

$$E_{in}(t) = E_0 e^{j\omega_0 t + j\beta x(t)} \tag{9}$$

where $\beta$ is the modulation index defined as $\pi V_p/V_\pi$. Here $V_\pi$ corresponds to the half-wave voltage of a phase modulator. We achieve field-to-field differentiation when the MZI-based differentiator is biased at the null point with a frequency-domain response of $j(\omega - \omega_0)$. The output signal in frequency domain can then be written as:

$$E_{out}(\omega) = E_{in}(\omega)H(\omega) = j(\omega - \omega_0)E_{in}(\omega) \tag{10}$$

Converting the output signal back into time domain through an inverse Fourier transform yield:

$$E_{out}(t) = \mathcal{F}^{-1}[E_{out}(\omega)] = \frac{dE_{in}(t)}{dt} - j\omega_0 E_{in}(t)$$
$$= j\beta \frac{dx(t)}{dt} E_0 e^{j\omega_0 t + j\beta x(t)} \tag{11}$$

Here since the output differentiation result is represented as optical field without a constant DC component, the PD picks up the output intensity which represents the square of the calculated derivative:

$$I_{out}(t) \propto \Re|E_{out}(t)|^2 = \Re\beta^2 E_0^2 \left[\frac{dx(t)}{dt}\right]^2 \tag{12}$$

As a result, all measured signals for field-to-field differentiation processes in this work are positive only.
On the other hand, we achieve field-to-intensity differentiation when biasing the MZI at the quadrature point. For simplicity, we assume the slope of the frequency response to be 1, in this case the output signal in frequency domain can be written as:

$$E_{out}(\omega) = E_{in}(\omega)H(\omega) = (\omega - \omega_n)E_{in}(\omega) \tag{13}$$

where $\omega_n$ is the null point of the MZI-based differentiator. Converting the output signal back into time domain by an inverse Fourier transform yields:

$$E_{out}(t) = \mathcal{F}^{-1}[E_{out}(\omega)] = -j\frac{dE_{in}(t)}{dt} - \omega_n E_{in}(t)$$
$$= \left[(\omega_0 - \omega_n) + \beta \frac{dx(t)}{dt}\right] E_{in}(t) \tag{14}$$

The output intensity captured by the PD is then:

$$I_{out}(t) \propto \Re|E_{out}(t)|^2 = \Re\left[(\omega_0 - \omega_n) + \beta \frac{dx(t)}{dt}\right]^2 E_0^2$$
$$= \Re E_0^2 \left\{(\omega_0 - \omega_n)^2 + \beta^2 \left[\frac{dx(t)}{dt}\right]^2 + 2\beta(\omega_0 - \omega_n)\frac{dx(t)}{dt}\right\} \tag{15}$$

where only the third term is effectively picked up by the AC-coupled PD under small signal modulation ($\beta \ll 1$), similar to the field-to-intensity integration case. Therefore, the expression can be simplified as:

$$I_{out}(t) \approx 2\beta\Re E_0^2(\omega_0 - \omega_n)\frac{dx(t)}{dt} \tag{16}$$

**Methodologies of the integration, differentiation and image edge-detection experiments:**
In our experiment, different analog signals are pre-encoded by an arbitrary waveform generator (AWG, Keysight M8199A) and loaded into the LN-based on-chip MWP system using the same high-speed photonic chip measurement setup as discussed before. Continuous wave optical carrier from the tunable laser is first amplified using an EDFA (Amonics, C-band) before sending into the chip thanks to the high power-handling capability of our LN devices, therefore minimizing the additional noise from EDFA when amplifying small signals. The output light of the LN chip is detected by a high-speed photodetector (Finisar XPDV412xR) and sent to a high-speed oscilloscope (Keysight UXR0592AP) for signal analysis. The frequency responses of the link components, including the AWG, cables, probes, adaptors and the oscilloscope, are calibrated and de-embedded up to 62 GHz. For the temporal integration experiment, the analog signals are combined with a DC bias voltage through a bias-tee to keep the MZI modulator at the quadrature point. The laser wavelength is fine tuned to align with the corresponding operating wavelengths of the integrators and differentiators. For UWB generation, the output spectrum is measured using an RF spectrum analyzer (R&S, FSW43) from 100 MHz to 15 GHz.

The photonic-assisted image edge detector is based on the field-to-field differentiator, which benefits from the direct intensity detection at the PD. The image information is encoded into a time-domain signal by serializing the pixels into a sequence of pulses and output at a highest sampling rate of 256 GSa/s using the AWG.

**Analog signal processing bandwidth analysis:**

More details are provided here for estimating the actual processing performances of the LN-based MWP system beyond the analog bandwidth of our oscilloscope, by inputting sawtooth signals at different repetition rates. The ideal Fourier transform of a sawtooth signal (Extended Data Fig. 1a) includes infinite numbers of frequency components at integer multiples of the fundamental frequency (or repetition rate). The MWP system processing bandwidth could therefore be inferred from the number of harmonics preserved in the final optical spectrum and their relative strengths. In our system, the AWG and electronic accessories (cables, adaptors, and probes) could support the generation and delivery to the chip of signals at up to 67 GHz, while the oscilloscope could only record the differentiated signal in time domain up to a 62-GHz bandwidth. Meanwhile the processing bandwidth of our MWP chip should in principle be beyond 70 GHz based on the EO modulation bandwidth measurement. In Extended Data Fig. 1b, we list the 3-dB bandwidths of these components for reference. Extended Data Fig. 1c-d demonstrates the temporal differentiation results recorded by oscilloscope and their corresponding optical spectra for sawtooth repetition rates of 20, 22, and 24 GHz, respectively. At 20 GHz, the temporal differentiation result shows good agreement with simulation result (red dashed) when considering only the first three harmonics, consistent with the measured optical spectrum showing three peaks at 20 GHz, 40 GHz, and 60 GHz. At 22 GHz, the temporal result shows substantial discrepancy (only one small lobe between the peaks) with the simulated result (two small lobes), since the oscilloscope is cut off at 62 GHz and cannot effectively pick up the third harmonic component at 66 GHz here. However, we can still clearly observe the three harmonics in the optical spectrum with relative strengths in line with the simulated values (red solid), from which we could infer the actual temporal performance of our differentiator by comparing the spectral responses with simulation. At an even higher repetition rate of 24 GHz, the third harmonic component is submerged into the noise floor since this 72-GHz component exceeds the analog bandwidth of the AWG and is not effectively generated in the first place. In Extended Data Fig. 1e, we summarize the measured the optical power roll-off of the third harmonic component (blue dots) for sawtooth signals at fundamental frequencies from 19 to 24 GHz, which is consistent with the simulation results (red line) at least up to 67 GHz. Further de-embedding the frequency responses of the test equipment and accessories allow us to infer the on-chip processing performance (red circles), which is in line with the simulated trend up to 70 GHz. Therefore, we conclude that our device can faithfully perform MWP processing tasks for signals with analog bandwidths up to ~ 70 GHz.

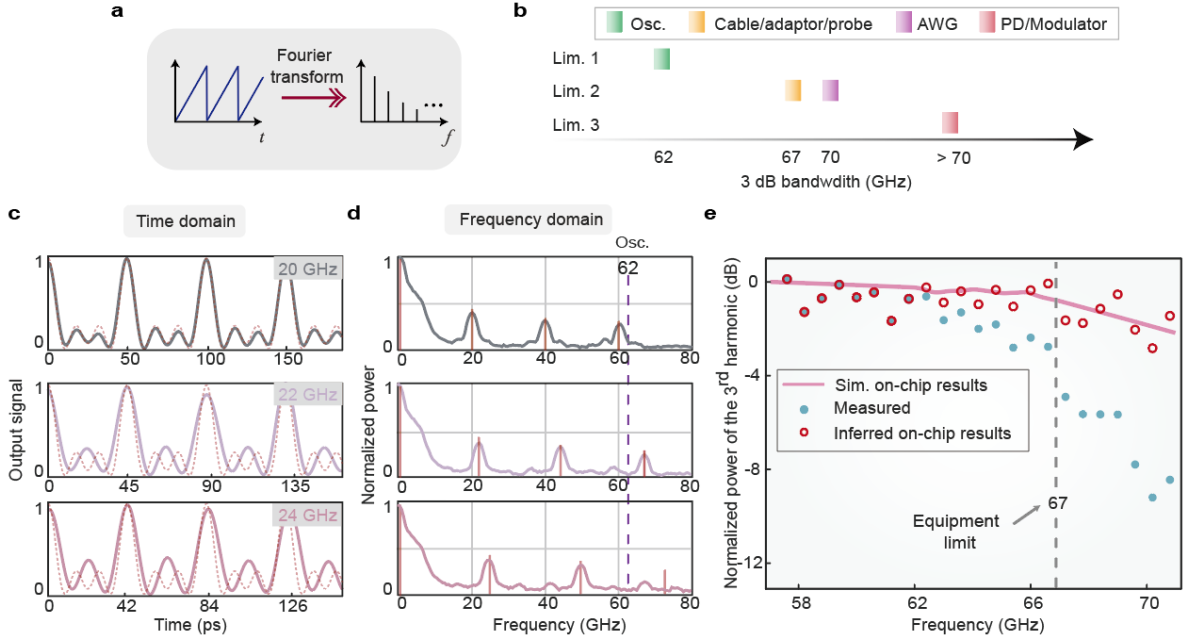

**Extended Data Fig. 1.** Analog signal processing bandwidth analysis based on sawtooth signal. **a** Time- and frequency- domain representations of sawtooth signal. **b** Analog bandwidth limitations of various components in our experiment. **c** Measured temporal differentiation results (solid) of sawtooth signals at fundamental frequencies of 20, 22, and 24 GHz, in comparison with simulated data (dashed) taking into consideration the first three harmonics. **d** Measured optical spectra of the differentiated signals, in comparison with the simulated results (red). **e** Measured (blue dots) optical power of the third-harmonic component of differentiated sawtooth signals at frequencies from 57 to 72 GHz, together with inferred on-chip performances after de-embedding equipment responses (red circles) and simulation results (red curve). Osc., oscilloscope. Lim., limitation. Sim., simulation.

**Image segmentation model and training process:**

More details are presented here for the principle of image segmentation model shown in Fig. 4c. We follow the network structure of DeepLabv3+ [55], which is composed of an encoder and a decoder. The former incorporates a ResNet101[56], Atrous Spatial Pyramid Pooling (ASPP) [55], and a 1×1 convolutional layer (conv) to extract deep features. The latter aims to recover the original image resolution and produce segmentation results. To optimize the segmentation model, we first collect simulated edge data for each medical image, which are concatenated together to feed into the encoder of the segmentation model, emphasizing the representations around melanoma boundaries. The deep features extracted from the encoder are then bilinearly up-sampled by a factor of 4 and concatenated with the corresponding low-level features from the ResNet101 backbone, which can preserve the detailed texture information, such as edges. These concatenated feature maps are further passed through a $3 \times 3$ convolutional layer, followed by a pixel-level classifier and a bilinear up-sampling operator, thereby deriving segmentation predictions. The obtained predictions are constrained by the dice loss [57]. By minimizing the dice loss, the optimized model has the ability to outline melanoma lesions given any test dermoscope images.

In implementation, we firstly construct a training dataset with 1000 dermoscope images [58], together with the corresponding simulated edge information and melanoma segmentation ground truths. Each dermoscope image is concatenated with its simulated edge data, and then the concatenated data are passed through the cascaded encoder and decoder of the segmentation model, obtaining segmentation prediction from the output of the decoder. The resulting prediction is supervised by the corresponding segmentation ground truth through minimizing the computed dice loss. Our method is implemented with the PyTorch library. The input images are uniformly resized to 250×250 for training. The stochastic gradient descent optimizer is adopted with an initial learning rate of $10^{-5}$ for the pre-trained encoder and $10^{-4}$ for the rest trainable parameters within the segmentation model with random initialization. Polynomial learning rate scheduling is adopted with the power of 0.9. We choose a batch size of 8 and the maximum epoch number of 200 to guarantee the convergence of training.

**Error analysis and performance evaluation:**

All error analyses in this work are performed by calculating the mean absolute error (MAE) between the experimentally measured results and the ideal ones within the differentiation/integration time. We model and analyze the errors mainly from three sources: *i*) The limited bandwidth of measurement devices. In our measurement setup, the oscilloscope does not faithfully capture the high-frequency components of the output signals, resulting in the distortion of output signal in time domain, as illustrated in Extended Data Fig. 2a in the case of an 8-ps Gaussian pulse. We expect the MAE to decrease from 5.8 % to 1.3 % if the ideal test equipment is used. *ii*) The possible drift of device operation points. Extended Data Fig. 2b demonstrates the simulated output result where the bias point for field-to-field differentiation drifts slightly by 3 MHz, which clearly exhibits an asymmetric doublet pulse for a 30-ps input Gaussian pulse. *iii*) The intrinsic processing bandwidths of the proposed integrator/differentiator devices, which are usually limited by the FSRs of the microring or MZI. As shown in Extended Data Fig. 2c-d, taking a short Gaussian pulse as an example (under the ideal test equipment), the pulse bandwidth is much broader than the FSRs of the ring-based integrator (80 GHz) or the MZI-based differentiator (75 GHz), as illustrated in the right insets, leading to clear distortion of the processing results.

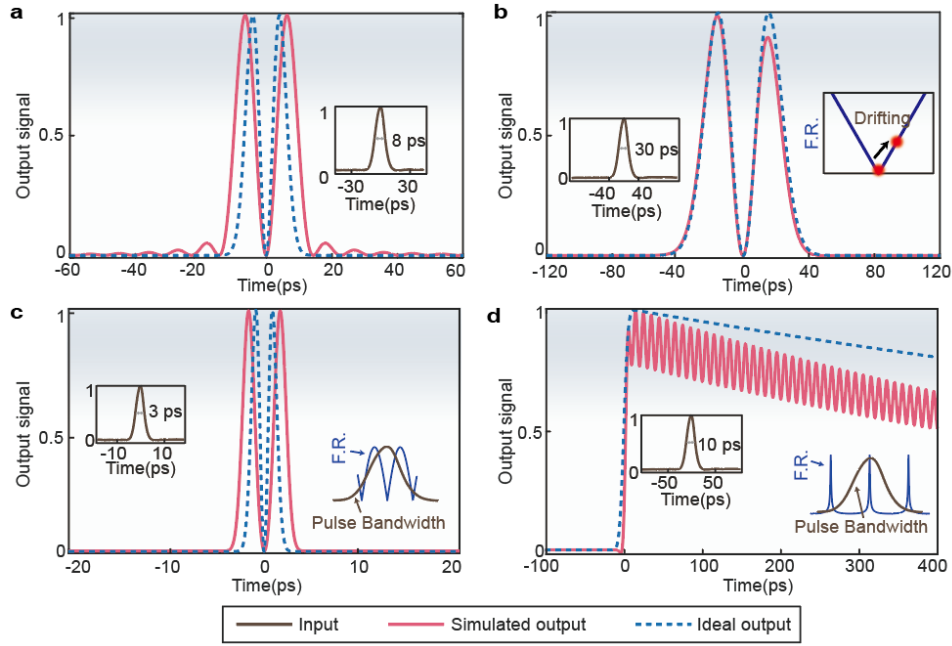

**Extended Data Fig. 2.** Error analyses based on simulated output signals with considerations of various practical limitations, including **a** limitation of equipment bandwidth, **b** drifting of device operation point, as well as the processing bandwidth limitations of the **c** MZI-based differentiator and **d** ring-based integrator. F.R., frequency response of the devices.

In real applications, the proposed high-speed microwave signal processor should support accurate signal processing within a broad bandwidth. To evaluate the error performance over a broad bandwidth, we take the ultrahigh-speed differentiation system as an example. The measured frequency response of the device is utilized for error analysis. The simulated MAE values of the functional devices (Extended Data Fig. 3) at different cut-off frequencies of Sinc pulses are analyzed firstly by assuming ideal measured equipment (red curve) and then by considering our current equipment bandwidths, consistent with our measured error performances (blue crosses and purple stars). Low computation errors of less than 4 % (dash line) could be maintained within broad bandwidth for the differentiator. The proposed MWP signal differentiators can accurately operate up to ~ 100 GHz assuming ideal test equipment.

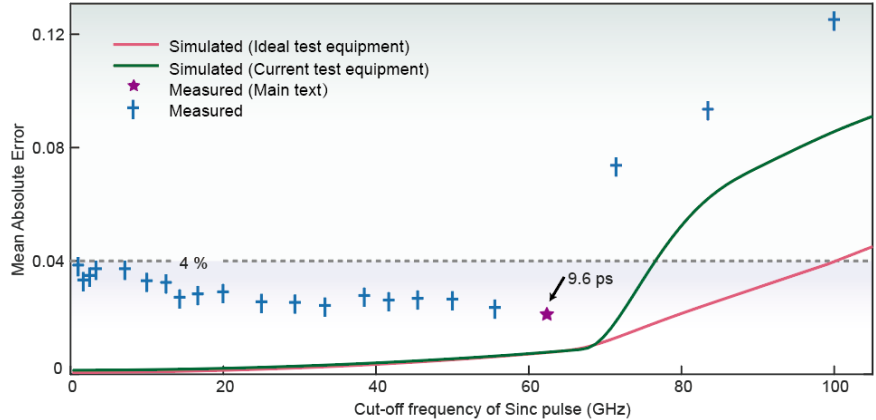

**Extended Data Fig. 3.** Simulated and measured mean absolute errors as functions of the cut-off frequency of Sinc pulse for the temporal differentiator.

**Performance comparison with electronic computer-based algorithms and hardware**

Extended Data Table 1 lists a detailed performance comparison between traditional electronics-based algorithms (including convolution-based and simple differentiation-based algorithms) and our photonic-assisted segmentation model. The performance metrics include raw lesion edge detection accuracy (from the edge detectors only, before DCNN), final segmentation accuracy (after DCNN), computation time, and energy consumption. The accuracies of lesion edge detection and segmentation are measured by dice coefficient. Extended Data Fig. 4 shows the raw edge detection results of an example lesion image using different processing methods, together with the ground truth. Our photonic edge detector shows a better raw edge detection accuracy (21 %) than those of both convolution-based (18.1 %) and differentiation-based (12.8 %) algorithms, mainly because it picks up less false-positive details inside the lesion region. The final image segmentation accuracies are above 95 % for all three methods but with drastically different processing time. For edge feature extraction of a 250×250-pixel image, our device consumes a total computation time of 244 ns, above three orders of magnitude faster than performing a traditional convolution algorithm on a generic personal computer (380 μs).

**Extended Data Table. 1. Performance comparison with traditional electronics-based algorithms**

|  | Differentiation algorithm [#] | Convolution algorithm [#] | This work |
|---|---|---|---|
| **Raw lesion edge detection** [*][†] | 12.8% | 18.1% | 21% |
| **Segmentation accuracy**[*] | 95.6% | 95.9% | 97.3% |
| **Computation time** | 120 μs | 380 μs | 0.244 μs |
| **Energy consumption** | 201 nJ | 1227 nJ | 3.76 nJ |

[*] Detection and segmentation accuracies are measured by dice coefficient.
[†] Raw detection accuracy right after edge detection algorithms, before entering DCNN.
[#] Electronic computer-based algorithms are processed by Intel Xeon(R) Gold 5215 CPU.

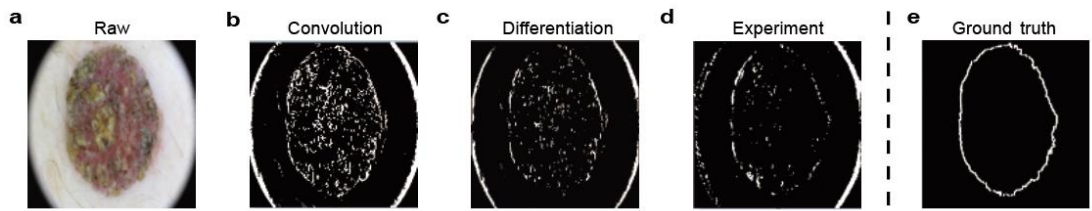

**Extended Data Fig. 4.** Edge-detection results of a representative melanoma lesion image. **a.** Raw image. **b-d**. Edge-extracted images using convolution (**b**) and simple differentiation (**c**) algorithms in electronic computers, and our photonic-assisted edge detector (**d**). **e**. Ground truth.

When estimating the energy consumption of electronics-based edge detection, we realize the challenges of quantifying the exact power consumption of DSP at the chip level, and only provide a lower-bound estimation by calculating the power consumed on the basic multiplication and addition operations of the edge-detection algorithms [59]. The energy consumption could be estimated as:

$$E_{ele} = T \cdot M \cdot [N \cdot E_M + (N-1) \cdot E_A] + (T-1) \cdot M \cdot E_A \quad (17)$$

where $T$ is the kernel number, which is typically 4 for a convolution algorithm and 2 for a differentiation algorithm. $M$ is the image pixel number, which is 250×250 in this case. $N$ is the kernel size, which is 9 for the convolution algorithm and 3 for the differentiation algorithm. We assume 7 nm CMOS node is used for DSP chips, with an energy consumption per multiplication operation $E_M$ ~ 486 fJ/b and an energy consumption per addition operation $E_A$ ~ 60.8 fJ/b [60]. Therefore, the estimated total energy consumption for convolution algorithm is ~ 1227 nJ and for differentiation algorithm is ~ 201 nJ.

For our photonic-assisted method, we estimate the comprehensive energy dissipation of our demonstrated MWP system by calculating the energy consumptions of the laser, EDFA, modulator as well as PD [37,59-61]. The equations used and values estimated are listed in Extended Data Table 2.

**Extended Data Table. 2. Estimated energy consumption of the photonic-assisted image edge detector**

| Component | Equation | Energy consumption |
|---|---|---|
| Laser | $E_{laser} = [(P_{laser} + P_{TEC})/\eta] \cdot t$ | 1.87 nJ |
| Modulator (on-chip) | $E_{mod} = (V_P^2/2R) \cdot t$ | 0.61 nJ |
| PD | $E_{PD} = \frac{\Re V_{bias} p_{rec}}{B} \cdot M$ | 0.44 nJ |
| EDFA | $E_{EDFA} = \frac{\lambda_s}{\lambda_p}[(P_{out}^s - P_{in}^s)/\eta] \cdot t$ | 0.84 nJ |
| Total | $E_{laser} + E_{mod} + E_{PD} + E_{EDFA}$ | 3.76 nJ |

Here $E_{laser}$ is the energy consumption of the pump laser, considering an optimized input optical power $P_{laser}$ of 1 mW and a thermo-electric cooler (TEC) module (for stabilizing the laser wavelength) power ~ 1.3 mW [62], $\eta$ is the wall-plug efficiency ~ 0.3, defined as the energy conversion efficiency from electrical power into optical power, $t$ ~ 244 ns is the processing time for the image edge detection task. Therefore, the consumed energy of $E_{laser}$ is 1.87 nJ. $E_{mod}$ is the energy consumption of the EO modulator, which intakes a driving peak voltage of $V_p$ = 500 mV (small-signal modulation) and consumes an average power consumption $\frac{V_p^2}{2R}$ of = 2.5 mW, where $R$ = 50 Ω is the load impedance of the modulator [37], leading to an $E_{mod}$ of ~ 0.61 nJ within the processing time. $E_{PD}$ is the energy consumption of the PD, which could be estimated as $\frac{\Re V_{bias} p_{rec}}{B}$ [61], where $\Re$ is the responsivity of the PD ~ 0.6 A/W, $V_{bias}$ is the PD bias voltage ~ 3 V, $p_{rec}$ is the received optical power ~ 1 mW, $B$ = 256 Gbit/s is the information bit rate, $M$ = 250×250 is the image pixel number, leading to $E_{PD}$ ~ 0.44 nJ. For the EDFA, the energy consumption could be estimated by $\frac{\lambda_s}{\lambda_p}(P_{out}^s - P_{in}^s)/\eta \cdot t$ [63], where $\lambda_s$ is signal wavelength ~ 1550 nm, $\lambda_p$ is the pump wavelength ~ 1480 nm, $P_{out}^s$ and $P_{in}^s$ are output signal power ~ 0 dBm and input signal power ~ -20 dBm respectively, $\eta$ is the wall-plug efficiency ~ 0.3, leading to $E_{EDFA}$ ~ 0.84 nJ. Therefore, the total energy consumption of our LN-based MWP system is estimated to be ~ 3.76 nJ.

**Performance comparison with previous literatures and other MWP platforms**

Extended Data Table 3 lists the detailed performance comparison among previously demonstrated all-optical processors, off-chip modulator-based MWP systems, an estimated all-Si integrated MWP system (since such demonstrations have not been reported), and our LN-based integrated MWP system. The performance metrics include MWP processing bandwidth, shortest processed input pulse length, highest demonstrated sampling rate of analog signals, and estimated energy consumption for image edge-detection tasks. For all-optical signal processors, the input pulses are generated by mode-lock lasers with pulse shapes limited to simple Gaussian or Gaussian-derived waveforms, which is hard to realize the complicated and arbitrary signal processing tasks demonstrated in this work. In other on-chip and off-chip MWP demonstrations, the processing bandwidths are generally limited by the EO modulator (whether off-the-shelf LN modulator or Si modulator), restricting the shortest processable input pulses to several tens of picoseconds. Furthermore, we estimate the energy consumption required in these previously demonstrated systems, assuming similar optical pump power, EDFA power, EO modulation depth, and PD performances, based on the methodologies and equations given in the above section and in Extended Data Table. 2. It should be noted that additional DC energy consumption is included for Si modulators due to the requirement of a reversed bias voltage [64]. As can be seen from Extended Data Table 3, our proposed LN-based MWP system, benefitting from the broadband, low-voltage, and linear EO responses, together with low optical loss and excellent scalability, exhibits significant breakthrough in terms of working bandwidth, processing speed and energy consumption compared with other MWP demonstrations.

**Extended Data Table. 3. Performance comparison with previous MWP demonstrations**

|  | MWP system bandwidth | Shortest processable input pulse | Compressed pulse by nonlinearity | Sampling rate of analog signals | Energy consumption for image edge detection tasks |
|---|---|---|---|---|---|
| **All-optical signal processor** | N/A | 33 ps [6] | N/A | N/A | N/A |
|  | N/A | 250 fs [28] | N/A | N/A | N/A |
|  | N/A | 7.5 ps [27] | N/A | N/A | N/A |
| **Off-chip modulator + on-chip signal processor** | 30-40 GHz (typ.) | 40 ps [21] | N/A | 65 GSa/s [21] | 22.05 nJ (est.) |
|  |  | 30 ps [20] | 18 ps [20] | N/A |  |
|  |  | 85 ps [8] | N/A | 65GSa/s [8] |  |
|  |  | N/A | 5.4 ps [26] | N/A |  |
|  |  | 290 ps [9] | N/A | N/A |  |
| **All-Si platform (est.)** | 33 GHz [3] | N/A | N/A | 50 GSa/s [3] | 128.65 nJ (est.) |
| **This work** | > 67 GHz | 9.6 ps | N/A | 256 GSa/s | 3.76 nJ |

typ., typical; est., estimated; N/A=Information not available or not applicable

**Funding.**

National Natural Science Foundation of China (61922092); Research Grants Council, University Grants Committee (CityU 11204820, CityU 21208219, N_CityU113/20, C1002-22Y); Croucher Foundation (9509005); Innovation and Technology Fund (ITS/226/21FP); City University of Hong Kong (9610402, 9610455).

**Acknowledgments.**

We thank Prof. Hon Ki Tsang for the use of high-speed measurement equipment. We thank the technical support of Mr. Chun Fai Yeung, Mr. Shun Yee Lao, Mr. C W Lai and Mr. Li Ho at HKUST, Nanosystem Fabrication Facility (NFF) for the stepper lithography and PECVD process. We thank Dr. Wing-Han Wong and Dr. Keeson Shum at CityU for their help in device fabrication and measurement.

**Author contributions.**

H.F. and C.W. conceived the idea. H.F. designed and fabricated the wafer with the help of Z.C., Y.Z., K.Z. and W.S. T.G. and X.G. performed the numerical simulations. H.F., T.G. and B.W. carried out the high-speed measurement and analyzed the data with the help of X.G., S.Z. and Y.Z. H.F. prepared the manuscript with contribution from all authors. C.H., Y.Y. and C.W. supervised the project. H.F. and T.G. contributed equally to this work.

**Conflict of interest.**

The authors declare no conflicts of interest.

**Data availability.**

Data underlying the results presented in this paper are not publicly available at this time but may be obtained from the authors upon reasonable request.